\documentclass[a4paper,11pt]{article}
\usepackage{jheppub}
\usepackage{natbib}
\usepackage{amsmath, amssymb, graphics, setspace} 
\usepackage{pdfpages}
\usepackage{color}
\usepackage{mathrsfs} 
\usepackage{multirow} 
\usepackage{bm}
\newcommand{\mathsym}[1]{{}}
\newcommand{\unicode}[1]{{}}

\usepackage{hyperref}
\hypersetup{
colorlinks=true,
pdfstartview=FitV,
linkcolor=blue,
citecolor=blue,
urlcolor=blue
}

\begin{document}
\title{Shear transport in far-from-equilibrium isotropization
of supersymmetric Yang-Mills plasma}
\author[a,h]{Shoucheng Wang,}
\author[b,c,d,e,*]{Song He}
\author[a,f,g,*]{Li Li}
 \affiliation[a]{Institute of Theoretical Physics,
Chinese Academy of Sciences, Beijing 100190, China}\affiliation[b]{Institute of Fundamental Physics and Quantum Technology,  Ningbo University, Ningbo, Zhejiang 315211, China}
\affiliation[c]{School of Physical Science and Technology, Ningbo University, Ningbo, 315211, China}
\affiliation[d]{Center for Theoretical Physics and College of Physics, Jilin University, Changchun 130012, People's Republic of China}
\affiliation[e]{Max Planck Institute for Gravitational Physics (Albert Einstein Institute), Am M\"uhlenberg 1, 14476 Golm, Germany}
\affiliation[f]{School of Fundamental Physics and Mathematical Sciences, Hangzhou Institute for Advanced Study, University of Chinese Academy of Sciences, Hangzhou 310024, China}
\affiliation[g]{School of Physical Sciences, University of Chinese Academy of Sciences, Beijing 100049, China}
\affiliation[h]{School of Science, Hunan Institute of Technology, Hengyang 421002, China}
\emailAdd{scwang@itp.ac.cn, hesong@nbu.edu.cn, liliphy@itp.ac.cn}

\let\oldthefootnote\thefootnote
\renewcommand{\thefootnote}{}
\footnotetext{*Corresponding author.}
\let\thefootnote\oldthefootnote

\abstract
{We holographically study the far-from-equilibrium isotropization dynamics of the strongly coupled $\mathcal{N}=4$ supersymmetric Yang-Mills plasma. The dual gravitational background is driven to be out of equilibrium and anisotropic by a time-dependent change in
boundary conditions. At late times, the system relaxes and asymptotically approaches a static configuration. The large
initial energy densities accelerate the isotropization significantly compared to the initial geometry corresponding to the supersymmetric Yang-Mills vacuum. We analyze shear transport during isotropization by directly computing the time-dependent stress tensor, which is now a nonlinear function of the shear rate.
The shear viscosity far from equilibrium displays much richer dynamics than its near-equilibrium counterpart. Moreover, we uncover that the equilibrium viscosity-to-entropy ratio at late times depends on the details of the quench function and the initial data, which could be due to a resummation of the hydrodynamic description. In particular, this ratio can be parametrically smaller than the Kovtun-Son-Starinets bound calculated from linear response theory.}

\maketitle
\section{Introduction}

The isotropization of non-equilibrium states in QCD and other non-Abelian quantum field theories has become a pivotal topic due to its broad relevance in heavy-ion collisions, early universe cosmology, and other domains. In the initial stages of relativistic heavy-ion collisions, the parton distribution is highly anisotropic, with transverse momenta significantly exceeding longitudinal momenta. As these partons interact and scatter, the stress tensor becomes isotropic, allowing the parton gas to transition into a quark-gluon plasma (QGP). This far-from-equilibrium matter formation after the initial interaction between colliding nuclei is complex, as different stages of the collision, embedded in a strongly coupled system, cannot be isolated experimentally~\cite{Heinz:2013th, Busza_2018}. Particularly challenging is the understanding of the initial state and pre-hydrodynamic stages.

A key insight from the past decade is that the apparent success of hydrodynamic modeling RHIC collisions~\cite{Heinz:2004pj} does not only imply fast isotropization of QGP, but also reflects the rapid decay of non-hydrodynamic modes—a process termed hydrodynamization—where the system enters a regime describable by viscous hydrodynamics long before achieving full isotropy~\cite{Busza_2018,Chesler:2009cy,Heller:2011ju,Heller:2013fn,Florkowski:2017olj,Kurkela:2018wud,Berges:2020fwq}. This distinction arises because hydrodynamic attractors, universal dynamical trajectories insensitive to initial conditions, allow effective hydrodynamic descriptions even when microscopic degrees of freedom remain far-from-equilibrium~\cite{Heller:2015dha,Soloviev:2021lhs,Jankowski:2023fdz}. Experimental data from RHIC and the LHC further support this picture, showing that hydrodynamic behavior emerges on timescales shorter than those required for complete isotropization~\cite{Gale:2013da, Parkkila:2021tqq}.

A proper toy model for studying the dynamics of a far-from-equilibrium, strongly coupled non-Abelian plasma in a controlled setting is strongly coupled $\mathcal{N}=4$ supersymmetric Yang-Mills theory. Holography can study the theory in the limit of large $N_c$ and large 't Hooft coupling $\lambda$. The creation and evolution of anisotropic, homogeneous strongly coupled $\mathcal{N}=4$ supersymmetric Yang-Mills plasma have been studied by solving the vacuum Einstein's equations in five-dimensional AdS spacetime, driven to anisotropy by a time-dependent change in boundary conditions~\cite{Chesler:2008hg}, see also~\cite{Chesler:2009cy, Heller:2012km, Heller:2013oxa}. These studies yielded short isotropization times for QGP, similar to estimates of thermalization times inferred from hydrodynamic modeling of RHIC collisions~\cite{Heinz:2004pj}. The anisotropic initial states relax via quasinormal mode decay and nonlinear interactions, with entropy production governed by the growth of apparent horizons~\cite{Romatschke:2017vte,Alho:2020gwl,Mondkar:2021qsf,Baggioli:2021tzr}. Crucially, the hydrodynamization process highlights how isotropization is intertwined with suppressing non-hydrodynamic excitations rather than mere local thermalization.  Additionally, holographic approaches provide valuable insights into heavy-ion collisions, with hydrodynamization times closely aligning with experimental data, demonstrating their effectiveness in early-stage collision modeling~\cite{Muller:2020ziz, Nijs:2023yab}.


Despite extensive research on isotropization dynamics in the holographic literature, shear viscosity in far-from-equilibrium isotropization has not yet been thoroughly explored. Viscosity is a fundamental property of liquid dynamics, characterizing a fluid's resistance to shear motion. For strongly interacting quantum field theories that admit gravity duals, holographic computations using linear response theory (the Kubo formula) have led to a conjectured viscosity bound, known as the Kovtun-Son-Starinets (KSS) bound~\cite{Kovtun:2004de}. When the deformation rate becomes large, the viscosity becomes a nonlinear function of the shear rate, producing many interesting and ubiquitous phenomena. In particular, the fluid enters a far-from-equilibrium state beyond the linearized hydrodynamics description. It is well known that the early-time dynamics of the QGP are characterized by strong spatial anisotropy. The apparent viscosity $\eta$ out of equilibrium is generally a function of both time and shear rate, displaying much richer dynamics than its near-equilibrium counterpart. In this regime, the gravitational solution becomes inherently time-dependent. A more appropriate and robust approach is to directly compute the time-dependent boundary stress tensor, a nonlinear shear rate function. More precisely, shear viscosity often manifests as anisotropy in the energy-momentum tensor's diagonal or off-diagonal components. For example, the off-diagonal approach can be found in using holographic tools in~\cite{Baggioli:2021tzr, Baggioli:2019mck}.\footnote{Another proposal for shear viscosity in out-of-equilibrium systems was suggested in~\cite{Wondrak:2020tzt} using linear response theory. The critical point is to compute the shear correlator in position space after turning on a shear perturbation and then perform a Wigner transformation to obtain the frequency corresponding to the relative time between source and response.} In this work, we consider the evolution of shear viscosity during the isotropization dynamics of far-from-equilibrium supersymmetric Yang-Mills plasma. We shall uncover rich dynamics of isotropization and shear transport. In particular, we will show that the late-time steady value of the viscosity-to-entropy ratio can differ from the KSS result.

The paper is structured as follows. Section~\ref{hmodel} presents the holographic setup for investigating isotropization dynamics driven to be anisotropic by a time-dependent change in boundary conditions. Section~\ref{result} discusses the far-from-equilibrium isotropization process, providing details on the evolution of the energy density and transverse and longitudinal pressures. Section~\ref{sec:shear} is devoted to the shear flows during isotropization, focusing on the evolution of shear viscosity and its deviations from its near-equilibrium counterpart. We summarize our findings and discuss future research directions in Section~\ref{sec:conc}.

\section{Holographic Setup}\label{hmodel}

The strongly coupled $\mathcal{N}=4$ supersymmetric Yang-Mills plasma is holographically described by a five-dimensional gravitational theory with a negative cosmological constant:
\begin{equation}
S = \frac{1}{2\kappa_{N}^2} \int d^5x \sqrt{-g} \left[ \mathcal{R}+\frac{12}{L^2}\right], \label{action}
\end{equation}
with $\mathcal{R}$ the Ricci scalar and $L$ the AdS radius that will be set to 1. The effective Newton constant $\kappa_{N}=2\pi/N_c$ with $N_c$ the rank of the $SU(N_c)$ gauge group. 

The holographic isotropization process is realized by placing the boundary theory in a time-dependent metric for a finite period of time~\cite{Chesler:2008hg}. The background metric $g_{\mu \nu}^{B}(x)$ that maintains spatial homogeneity, $O(2)$ rotational invariance, and a constant spatial volume element is given by
\begin{equation}
ds^{2} = -dt^{2} + e^{B_{0}(t)} d\bm{x}_{\perp}^{2} + e^{-2B_{0}(t)} dx_{\parallel}^{2}\,, \label{bgmetric}
\end{equation}
where $\bm{x}_{\perp} \equiv \{x_{1}, x_{2}\}$. The function $B_{0}(t)$ is chosen to be~\cite{Chesler:2008hg}:
\begin{equation}
\label{eqb0}
B_{0}(t) = \frac{1}{2} c \left[ 1 - \tanh{\left( \frac{t}{\tau} \right)} \right]\,,
\end{equation}
where $c$ represents a rescaling of transverse lengths relative to the longitudinal direction over a timescale of order $\tau$. This time-dependent boundary metric propagates gravitational radiation from the AdS boundary into the bulk. At late times, as the boundary geometry $g_{\mu \nu}^{B}$ becomes static, the bulk geometry relaxes and asymptotically approaches a static configuration. Without loss of generality, all quantities are expressed in units where $\tau = 1$.

The corresponding bulk configuration is given as follows
\begin{equation}
ds^{2} = -A(t,r) dt^{2} + \Sigma(t,r)^{2} \left[ e^{B(t,r)} d\bm{x}_{\perp}^{2} + e^{-2B(t,r)} dx_{\parallel}^{2} \right] + 2 dr dt\,,
\label{eq:metric}
\end{equation}
where the AdS boundary is located at $r \rightarrow \infty$, at which point the coordinate $t$ coincides with the boundary time of~\eqref{bgmetric}. The infalling radial null geodesics have constant values of $(t, \bm{x}_{\perp}, \bm{x}_{\parallel})$. The apparent horizon, located at $r = r_h$, for the background~\eqref{eq:metric} is determined by
\begin{equation}
    (\partial_t\Sigma+\frac{1}{2} A \partial_r\Sigma)|_{r_h} = 0 \label{dsigmaequal0}\,,
\end{equation}
which is an outermost marginally trapped surface.

The functions $(A, B, \Sigma)$ are determined  by solving the Einstein’s equations
\begin{equation}
\mathcal{R}_{\mu\nu}-\frac{1}{2}
\mathcal{R} g_{\mu\nu} - 6 g_{\mu\nu}=0\,,
\end{equation}
with the explicit form of equations of motion given as 
\begin{subequations}\label{motionequations}
\begin{align}
0 &= \Sigma'' + \frac{1}{2} \Sigma B'^{2}\,, \label{eqmotion1} \\
0 &= d_{+} \Sigma' -2 \Sigma + \frac{2}{\Sigma} d_{+} \Sigma \Sigma'\,, \label{eqmotion2} \\
0 &= d_{+} B' + \frac{3}{2 \Sigma} d_{+} B \Sigma' + \frac{3}{2 \Sigma} d_{+} \Sigma B'\,, \label{eqmotion3} \\
0 &= A'' +4 + 3 d_{+} B B' - \frac{12}{\Sigma^{2}} d_{+} \Sigma \Sigma'\,, \label{eqmotion5} \\
0 &= d_{+} (d_{+} \Sigma) + \frac{1}{2} d_{+} B^{2} \Sigma - \frac{1}{2} d_{+} \Sigma A'\,, \label{constrain}
\end{align}
\end{subequations}
where (\ref{constrain}) is the constraint equation. For any function $f(r,t)$ we have defined
\begin{equation}
f'\equiv \partial_{r} f,\quad \dot{f}\equiv \partial_{t} f,\quad d_{+}f\equiv  \partial_{t}f+\frac{1}{2} A\partial_{r} f\,,
\end{equation}
with prime denoting the derivative with respect to radial coordinate $r$ and dot denoting the derivative with respect to time $t$. Those equations are subjected to boundary conditions imposed at both the inferior boundary (\emph{e.g.} the apparent horizon $r_{h}$) and the AdS boundary $r \to \infty$. 

The conditions required for the boundary metric to match (\ref{bgmetric}) are
\begin{equation}
\begin{aligned}\label{bdcondition}
&\lim_{r \to \infty} \frac{\Sigma(r,t)}{r} = 1\,, \\
&\lim_{r \to \infty} B(r,t) = B_{0}(t)\,, \\
&\lim_{r \to \infty} \frac{A(r,t)}{r^{2}} = 1\,.
\end{aligned}
\end{equation}
Moreover, the form of the metric~\eqref{eq:metric} allows the residual diffeomorphism $r\rightarrow f(t)$ with $f(t)$ an arbitrary function. We shall fix the residual diffeomorphism invariance by
demanding $\lim_{r \to \infty}[A(r,t)-r^2]/r=0$.\footnote{Another way is to put the apparent horizon $r_h$ at a fixed radial position, which makes the computational domain a simple rectangular region with $r_h<r<\infty$. In practice, the former results in a more stable numerical scheme for solving the equations of motion.}
Then, near the AdS boundary, we have the following asymptotic expansion.
\begin{equation}\label{uvexpand}
\begin{split}
A(r,t) &= r^2-\frac{5}{4} \dot{B}_{0}(t){}^2+\frac{a_{4}(t)}{r^2} +\frac{\ln (r) \left(-3 \dot{B}_{0}(t){}^4+2 \dddot{B}_{0}(t) \dot{B}_{0}(t)-\ddot{B}_{0}(t){}^2\right)}{8r^2} + \mathcal{O}\left(\frac{\ln(r)}{r^3}\right), \\
\Sigma(r,t) &= r - \frac{\dot{B}_{0}(t)^2}{4 r} - \frac{\dot{B}_{0}(t)\ddot{B}_{0}(t)}{12 r^2}  + \mathcal{O}\left(\frac{\ln(r)}{r^3}\right), \\
B(r,t) &= B_{0}(t) + \frac{\dot{B}_{0}(t)}{r} + \frac{\ddot{B}_{0}(t)}{4 r^2}\quad + \frac{-\dddot{B}_{0}(t) + 5\dot{B}_{0}(t)^3}{12 r^3} + \frac{b_{4}(t)}{r^4} \\
&\quad + \frac{\ln(r)}{16 r^4}\left(\ddddot{B}_{0}(t) - 6 \dot{B}_{0}(t)^2 \ddot{B}_{0}(t)\right) + \mathcal{O}\left(\frac{\ln(r)}{r^5}\right)\,, 
\end{split}
\end{equation}
together with a constraint relation 
\begin{equation}\label{ward_ceq}
\begin{split}
\dot{a}_{4}(t)=&4 b_4(t) \dot{B}_{0}(t)+\frac{17}{48}\dot{B}_{0}(t) \ddddot{B}_{0}(t)-\frac{5}{2}\dot{B}_{0}(t){}^3 \ddot{B}_{0}(t)-\frac{1}{8} \dddot{B}_{0}(t) \ddot{B}_{0}(t)\,.
\end{split}
\end{equation}

The stress tensor of the supersymmetric Yang-Mills theory can be obtained by holographic renormalization. One has~\cite{Chesler:2008hg}
\begin{equation}
{T^\mu}_\nu=\text{diag}(-\mathcal{E}, \mathcal{P}_\perp, \mathcal{P}_\perp\,, \mathcal{P}_\parallel) \,,   
\label{stresstensor}
\end{equation}
with
\begin{equation}\label{eqE}
\begin{split}
\kappa_N^2\mathcal{E} &=- \frac{3}{2} a_4(t)-\frac{1}{128} \left(14 \ddot{B}_{0}(t)^2+3 \dot{B}_{0}(t)^4-4 \dddot{B}_{0}(t) \dot{B}_{0}(t)\right)\,,
\end{split}
\end{equation}
\begin{equation}
\begin{split}
\kappa_N^2\mathcal{P}_\perp &=-\frac{1}{2} a_4(t)+2 b_4(t)+\frac{1}{384} \left(64 \ddddot{B}_{0}(t)+10 \ddot{B}_{0}(t)^2+21 \dot{B}_{0}(t)^4+4 \dddot{B}_{0}(t) \dot{B}_{0}(t)\right.\\&\left.-468 \dot{B}_{0}(t)^2 \ddot{B}_{0}(t)\right)\,,
\end{split}
\end{equation}
\begin{equation}\label{eqPT}
\begin{split}
\kappa_N^2\mathcal{P}_\parallel &= -\frac{1}{2} a_4(t)-4 b_4(t)+\frac{1}{384} \left(-128 \ddddot{B}_{0}(t)+10 \ddot{B}_{0}(t)^2+21 \dot{B}_{0}(t)^4+4 \dddot{B}_{0}(t) \dot{B}_{0}(t)\right.\\&\left.+936 \dot{B}_{0}(t)^2 \ddot{B}_{0}(t)\right)\,.
\end{split}
\end{equation}
\begin{figure}
    \centering
    \includegraphics[scale=0.8]{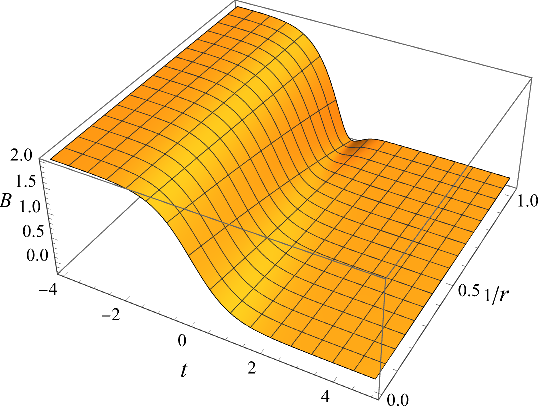}
    \includegraphics[scale=0.8]{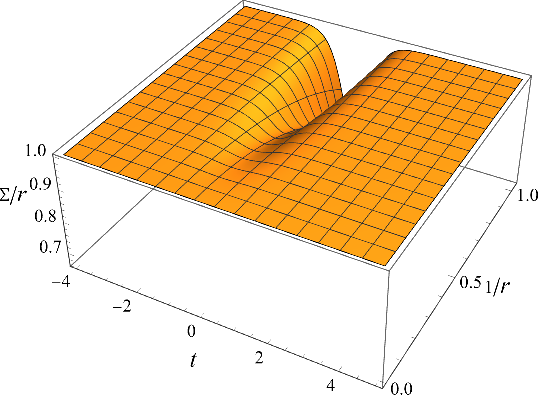}
    \caption{Illustration of the bulk configuration $(B, \Sigma/r)$ during far-from-equilibrium isotropization for $c=2$. In our coordinate system, the AdS boundary is located at $r\rightarrow\infty$ at which $B=B_0(t)$ is a boundary condition. Near $t=0$, the geometry exhibits significant anisotropy due to the quenching effect via~\eqref{eqb0}.}
    \label{fig:bulk}
\end{figure}
We show the fully non-linear evolution of the bulk dynamics in Figure~\ref{fig:bulk}. The bulk geometry starts in an equilibrium state at early times. As the dynamic boundary quench~\eqref{eqb0} is activated, the system progressively deviates from equilibrium, reaching a maximum deviation at $t = 0$. As the quench diminishes, the system relaxes and returns to thermal equilibrium at late times. 

\begin{figure}
    \centering
    \includegraphics[scale=1.]{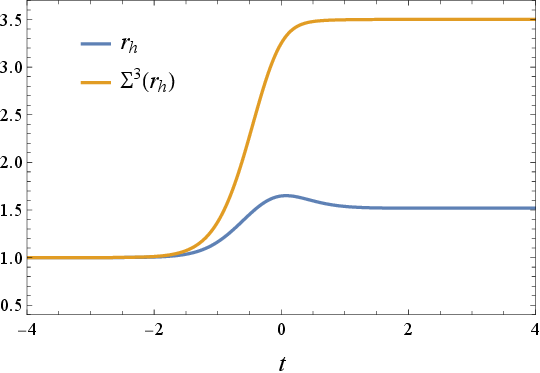}
    \caption{The evolution of the location of the apparent horizon $r_h$ and the apparent horizon area $\Sigma^3(r_h)$. The area of the apparent horizon increases monotonically as time evolves.}
    \label{fig:AP}
\end{figure}

The horizon position is time-independent in our coordinate system. The location of the apparent horizon $r_h$ and the apparent horizon area $\Sigma^3(r_h)$ are depicted in Figure~\ref{fig:AP}, for $c=2$. While the former (blue line), as a function of time, is non-monotonic, the area of the apparent horizon increases monotonically as time evolves. The growth of the area of the apparent horizon occurs nearly at intervals during which the boundary geometry changes rapidly. 

\section{Far-from-equilibrium isotropization}\label{result}

\begin{figure}
    \centering
    \includegraphics[scale=1.0]{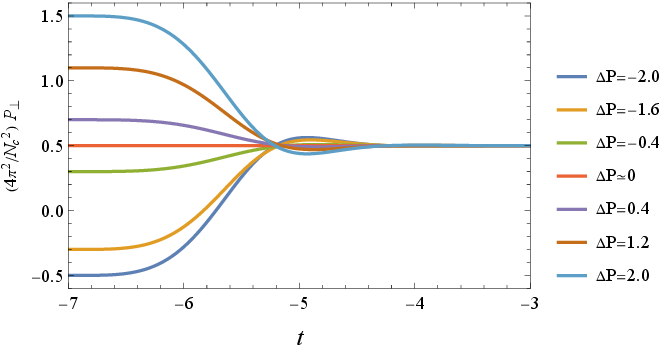}
    \caption{Transverse pressure ($P_{\perp}$) for various initial pressure anisotropy $\Delta P=\frac{P_{\perp}(t_0)-P_{\parallel}(t_0)}{\mathcal{E}(t_0)}$ with the same initial energy density. The pressure anisotropy increases as $b_4(t_0)$ is increased. We choose $a_4(t_0)=-1$ with $t_0=-7$ and $c=2$ for the dynamic boundary~\eqref{eqb0}.}
    \label{fig:b4}
\end{figure}

The initial conditions are characterized by the energy density and pressure anisotropy at the initial time $t_0\ll -1$, specifically $a_4(t_0)$ and $b_4(t_0)$ as defined in~\eqref{eqE}-\eqref{eqPT}. Figure~\ref{fig:b4} presents the numerical solutions for a range of initial conditions. The evolution exhibits a transient regime highly sensitive to the initial value of $b_4$ in~\eqref{uvexpand}. During this regime, the system's dynamics are dominated by UV microscopic details, resulting in nonspecific trends. However, our numerical analysis indicates that arbitrary initial data quickly converge towards an attractor driven by the decay of non-hydrodynamic modes. When $B_0(t)$ is nearly constant at early times, the pressure anisotropy originating from the initial data $b_4(t_0)$ decays rapidly. Simultaneously, from~\eqref{ward_ceq}, we observe that the energy density set by $a_4(t_0)$ remains approximately constant as $\dot{\mathcal{E}}(t) \approx 0$ during early times. Once the system reaches the attractor, arbitrary initial conditions are smoothed out via the decay of non-hydrodynamic modes. This analysis further suggests that the isotropization dynamics induced by the boundary quench~\eqref{eqb0} are largely insensitive to the initial pressure anisotropy.

\begin{figure}
    \centering
    \includegraphics[scale=0.8]{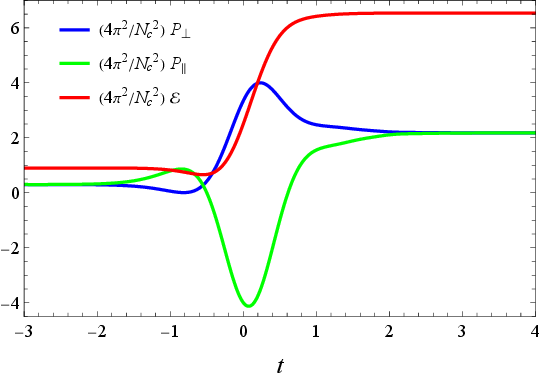}
    \includegraphics[scale=0.8]{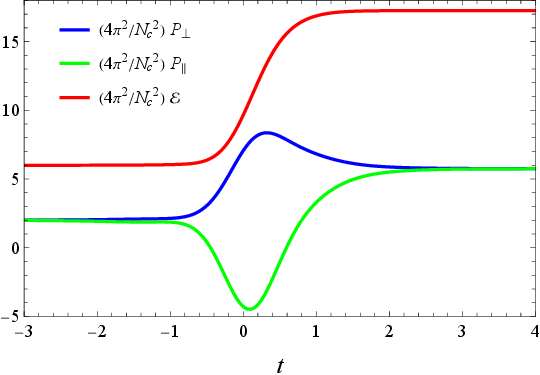}
    \caption{Time evolution of energy density $\mathcal{E}$, transverse pressure ($P_{\perp}$), and longitudinal pressure ($P_{\parallel}$) for $c=2$. Left panel corresponds to a small initial energy density ($a_4(t_0)=-0.6$) and right panel has a large one ($a_4(t_0)=-4$). Following a brief period of anisotropic geometry, all quantities converge to equilibrium late. It is manifest that the energy density appears to be driven below the ground state energy density in the first moments of the quench.}
    \label{fig:Tuv}
\end{figure}

Figure~\ref{fig:Tuv} shows the evolution of energy density, transverse pressure, and longitudinal pressure under the dynamic boundary conditions~\eqref{eqb0} with $c=2$. The system evolves from an initial state with finite positive values for $(\mathcal{E}, \mathcal{P}_\perp, \mathcal{P}_\parallel)$, initially satisfying $\mathcal{P}_\perp \approx \mathcal{P}_\parallel$. Subsequently, anisotropy develops ($\mathcal{P}_\perp \neq \mathcal{P}_\parallel$) due to the boundary quench. Non-monotonic behavior is observed when the boundary changes rapidly near $t=0$. In the left panel, with a small initial energy density, the energy density briefly dips around $t=-1$ before rising to a higher plateau. This feature corresponds to a negative apparent viscosity, as discussed in Section~\ref{sec:shear}. The dip vanishes with a higher initial energy density (right panel), leading to a monotonic increase before reaching equilibrium at late times. The thermal equilibrium state at late times is given by

\begin{equation}
T^{\mu\nu}_{eq}  =\frac{\pi^2N_c^2T_{eq}^4}{8} \text{diag}(3, 1, 1,1) \,,
\end{equation}
where $T_{eq}$ represents the final equilibrium temperature, which increases with the initial energy density $\mathcal{E}_0$. We explored different values of the amplitude $c$ and found qualitatively similar behavior during and after the quench.

\begin{figure}
    \centering
    \includegraphics[scale=0.8]{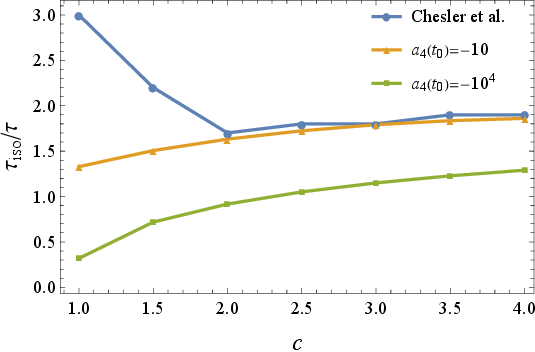}
    \includegraphics[scale=0.8]
    {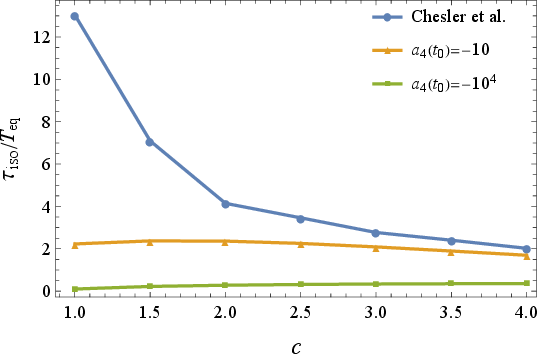}
    \caption{Isotropization time as a function of quench strength $c$ for different initial data. The isotropization time $\tau_{{iso}}$ is measured in units of $\tau$ (left) and equilibrium temperature $T_{eq}$ (right), respectively. Our results are compared with the isotropization time from Chesler et al.~\cite{Chesler:2008hg} where $a_4(t_0)\approx0$. The large initial energy densities accelerate the isotropization significantly.}
    \label{fig:tau}
\end{figure}

Following~\cite{Chesler:2008hg}, the isotropization time, $\tau_{\text{iso}}$, is defined as the time at which both transverse and longitudinal pressures reach within 10\% of their final equilibrium values. This timescale was argued to provide an upper bound on the isotropization times associated solely with plasma dynamics~\cite{Chesler:2008hg}. It was found that for the initial geometry corresponding to the vacuum state of supersymmetric Yang-Mills theory, \emph{i.e.} $\mathcal{E}_0 = \mathcal{P}_{\perp 0} = \mathcal{P}_{\parallel 0} = 0$, we have $\tau_{{iso}} \approx 2 \tau$ for $|c| > 2$ and $\tau_{{iso}} \approx 0.7 / T_{{eq}}$ for $|c| < 2$. To investigate whether isotropization is sensitive to the choice of the initial state, we compute $\tau_{{iso}}$ for different initial energy densities, characterized by an initial value of $a_4(t_0)$ (see Figure~\ref{fig:tau}). 
In contrast to the benchmark scenario considered in~\cite{Chesler:2008hg}, the behavior of $\tau_{{iso}}$ shows sensitivity to the initial energy density, $\mathcal{E}_0$. For $|c| > 2$, $\tau_{{iso}}/\tau$ appears to approach a constant, although the magnitude of this constant depends on $\mathcal{E}_0$. As depicted in the left panel of Figure~\ref{fig:tau}, this constant decreases with increasing $\mathcal{E}_0$. The feature $\tau_{{iso}} \approx 0.7 / T_{{eq}}$ for $|c| < 2$, reported in~\cite{Chesler:2008hg}, breaks down completely for higher values of $\mathcal{E}_0$ (orange and green curves in the right panel of Figure~\ref{fig:tau}). 
Therefore, compared to the vacuum state case of~\cite{Chesler:2008hg}, the large initial energy densities significantly accelerate isotropization.

Regarding equilibria, it is suggested that 
without fine-tuning (see however~\cite{Carballo:2024kbk}), the isotropization time is at most of the order of the background temperature. Unfortunately, since it is far from equilibrium, initial temperature is a poor measure since one can quickly pump more energy into the system. In practice, one can consider the final effective temperature $T_{eq}$ as the relevant scale that bounds thermalization dynamics. As shown in the right panel of Figure~\ref{fig:tau}, as $\mathcal{E}_0$ increases, $\tau_{{iso}}/ T_{{eq}}$ can decrease to near zero, instead of an order one value. This highlights the requirement to introduce a more reasonable isotropization time.

\section{Shear flows in far-from-equilibrium isotropization dynamics}\label{sec:shear}

The early-time dynamics of the system are characterized by strong spatial anisotropy. In this section, we examine the evolution of shear viscosity during the isotropization process in a far-from-equilibrium supersymmetric Yang-Mills plasma. We shall show to what extent strong time dependence and far-from-equilibrium physics affect shear transport in the supersymmetric Yang-Mills plasma.

\subsection{Definition of strain and stress}
As is evident from~\eqref{bgmetric}, the metric $g_{\mu \nu}^{B}$ of the boundary field theory is no longer flat, and additional care is required to define the boundary shear stress and shear strain. Following the approach of~\cite{Baggioli:2021tzr}, we define an apparent viscosity, $\eta$, in response to finite deformations. The geometry is decomposed into a fluid moving orthogonally to a homogeneous spatial hypersurface. In this context, the unit normal vector to the spatial hypersurface is identified with the timelike fluid 4-velocity, $u^\mu$, satisfying $g_{\mu \nu}^{B} u^\mu u^\nu = -1$. The induced metric on the spatial hypersurface is given by
\begin{equation}
    \gamma_{\mu \nu} = g_{\mu \nu}^{B}+u_\mu u_\nu\,,
\end{equation}
This metric also serves as a projection tensor. We then obtain the extrinsic curvature of the spatial hypersurface:
\begin{equation}
     K_{\mu \nu}^{B} = \gamma _{\mu}^{\sigma }\gamma _{\nu}^{\rho }\nabla_\rho u_\sigma=\omega_{\mu\nu}+\frac{1}{3}\Theta\gamma_{\mu \nu}+\sigma_{\mu \nu}\,.
\end{equation}
This quantity characterizes the variation in the 4-velocity direction across the spatial surface. Here, the antisymmetric part, $\omega_{\mu \nu} = K_{[\mu \nu]}^{B}$, is the vorticity tensor; the trace, $\Theta = g^{\mu \nu}_{B} K^B_{\mu \nu}$, represents the expansion scalar; and the symmetric traceless part is the shear tensor, $\sigma_{\mu \nu}$. For the metric in~\eqref{bgmetric}, the fluid velocity is given by $u^\mu = (u^t, u^{x_1}, u^{x_2}, u^{x_\parallel}) = (1, 0, 0, 0)$, from which we find $\omega_{\mu \nu} = \Theta = 0$, and the non-vanishing components of the shear tensor are

\begin{equation}
\sigma_{\perp\perp}=\frac{1}{2}e^{B_0(t)}\dot{B}_0(t), \qquad \sigma_{\parallel\parallel}=-e^{-2B_0(t)}\dot{B}_0(t)\,.
\end{equation}

On the other hand, the energy-momentum tensor, $T_{\mu \nu}$, for a relativistic fluid reads
\begin{equation}
     T_{\mu \nu}=\mathcal{E} u_\mu u_\nu+p\gamma_{\mu \nu}+2q(_\mu u_\nu)+\pi_{\mu \nu},
\end{equation}
where $\mathcal{E}=u^\mu u^\nu T_{\mu\nu}$, $p=\gamma^{\mu\nu}T_{\mu\nu}/3$ and $q_\mu=-{\gamma_{\mu}}^{\nu }u^\sigma T_{\nu \sigma}$ are the energy density, the pressure and the heat conduction vector. The dissipative transverse part $\pi_{\mu\nu}$ is the anisotropic stress tensor. The viscosity coefficient $\eta$ is introduced as
\begin{equation}\label{myeta0}
    \pi_{\mu \nu} = -2 \eta \, \sigma_{\mu \nu}\,.
\end{equation}
The above definition of $\eta$ accounts for nonlinear and non-hydrodynamic mode contributions~\cite{Romatschke:2017vte}. This effective viscosity has been adopted in viscous cosmology for early- and late-time universe (see~\cite{Brevik:2017msy} for a review).

Substituting~\eqref{myeta0} into the energy-momentum tensor of~\eqref{stresstensor}, we obtain
\begin{equation}\label{eta}
    \eta = \frac{\mathcal{P}_\parallel - \mathcal{P}_\perp}{3 \dot{B}_0(t)}\,,
\end{equation}
from which it follows that $\dot{B}_0(t)$ acts as the shear rate. Note that the mechanical deformations of the viscoelastic medium described by~\eqref{bgmetric} are characterized by $B_0(t)$. As shown in Figure~\ref{fig:Tuv}, $B_0(t)$ induces pressure anisotropy, \emph{i.e.} $\mathcal{P}_\parallel \neq \mathcal{P}_\perp$. At later times, when the shear rate $\dot{B}_0(t)$ approaches zero, the system relaxes and becomes isotropic asymptotically. Furthermore, regardless of the specific shear deformation applied, the behavior of the energy density is found to be
\begin{equation}\label{dotE}
    \dot{\mathcal{E}}(t) = 3 \eta \, \dot{B}_0(t)^2\,,
\end{equation}
using the constraint equation~\eqref{ward_ceq}.

The structure of~\eqref{dotE}, in particular, the perfecter, can be understood as follows. In our current coordinate system, the anisotropic stress tensor reads
\begin{equation}
    {\pi^\mu}_\nu = \left[\begin{matrix}
    \mathcal{P}_\perp - p & 0 & 0 \\
    0 & \mathcal{P}_\perp - p & 0 \\
    0 & 0 & \mathcal{P}_\parallel - p
    \end{matrix}\right] = \left[\begin{matrix}
    -\sigma & 0 & 0 \\
    0 & -\sigma & 0 \\
    0 & 0 & 2\sigma
    \end{matrix}\right] \,,
\end{equation}
where the isotropic pressure is $p = (2\mathcal{P}_\perp + \mathcal{P}_\parallel) / 3$, and the shear deformation is defined as $\sigma = (\mathcal{P}_\parallel - \mathcal{P}_\perp) / 3$. By performing a coordinate transformation, we have
\begin{equation}
    \begin{split}
        x &= \frac{1}{\sqrt{6}}(-\sqrt{3} x_1 - x_2 + \sqrt{2} x_\parallel)\,, \\
        y &= \frac{1}{\sqrt{3}}(\sqrt{2} x_2 + x_\parallel)\,, \\
        z &= \frac{1}{\sqrt{6}}(\sqrt{3} x_1 - x_2 + \sqrt{2} x_\parallel)\,,
    \end{split}
\end{equation}
such that the anisotropic stress tensor and the corresponding shear tensor in the new coordinate system are given by
\begin{equation}
    {\pi'^\mu}_\nu = \left[\begin{matrix}
    0 & \sigma & \sigma \\
    \sigma & 0 & \sigma \\
    \sigma & \sigma & 0
    \end{matrix}\right], \quad
    {\sigma'^\mu}_\nu = -\frac{\dot{B}_0}{2} \left[\begin{matrix}
    0 & 1 & 1 \\
    1 & 0 & 1 \\
    1 & 1 & 0
    \end{matrix}\right] \,,
\end{equation}
associated with the background metric
\begin{equation}\label{newbk}
    ds^2 = -dt^2 + \alpha(t)^2 (dx^2 + dy^2 + dz^2) + 2 \beta(t)(dx \, dy + dy \, dz + dz \, dx)\,,
\end{equation}
where $\alpha(t)^2 = \frac{e^{-2B_0(t)} + 2e^{B_0(t)}}{3}$ and $\beta(t) = \frac{e^{-2B_0(t)} - e^{B_0(t)}}{3}$. One can see that all spatial off-diagonal terms are equal. Moreover, we can easily check that
\begin{equation}\label{dfeta}
    \sigma = \eta \, \dot{B}_0\,,
\end{equation}
using~\eqref{eta}. Thus, shear deformations are carried out with the same shear rate, $\dot{B}_0$, in all three spatial directions. Note that~\eqref{dfeta} is the standard form of Newton's law of viscosity and is what is measured in the lab. 

Furthermore, from~\eqref{dotE}, we find that the energy density follows the evolution law of a viscoelastic system under deformation:
\begin{equation}\label{Etest}
    \Delta \mathcal{E}(t) \equiv \mathcal{E}(t) - \mathcal{E}_0 = 3 \int_{t_0}^t \dot{B}_0(\tau) \sigma(\tau) \, d\tau = 3 \int_{t_0}^t \eta(\tau) \, \dot{B}_0(\tau)^2 \, d\tau\,,
\end{equation}
where the subscript indicates the initial configuration upon which the shear deformation is applied. Therefore, the apparent shear viscosity quantifies the rate of energy change in response to the shear rate.

\subsection{Shear transport in far from equilibrium}
For non-equilibrium cases, a more appropriate definition for the entropy density is through the area element of the apparent horizon pulled back to the boundary along $t= const.$ infalling null geodesics~\cite{Hubeny:2007xt, Baggioli:2021tzr}.
\begin{equation}
\mathcal{S}(t)=\frac{2\pi\Sigma(t,r_h)^3}{\kappa_N^2}\,.    
\end{equation}
Then one obtains the entropy current density $s^\mu=\mathcal{S} u^\mu$, for which the entropy production is given by
\begin{equation}\label{eq:ds}
\nabla_\mu s^\mu=\frac{2\pi}{\kappa_N^2}\frac{\partial}{\partial t}\Sigma(t, r_h)^3\,.  
\end{equation}
Therefore, the non-negativity of entropy production is equivalent to the monotonic increase of the area of the apparent horizon. We find that the entropy production~\eqref{eq:ds} is always positive in all cases we consider, see \emph{e.g.} Figure~\ref{fig:AP}. Moreover, after the period at which the boundary geometry changes rapidly, the value of $\mathcal{S}$ quickly saturates to the thermal entropy of the final equilibrium state.

\begin{figure}[h]
    \centering
    \includegraphics[scale=1.0]{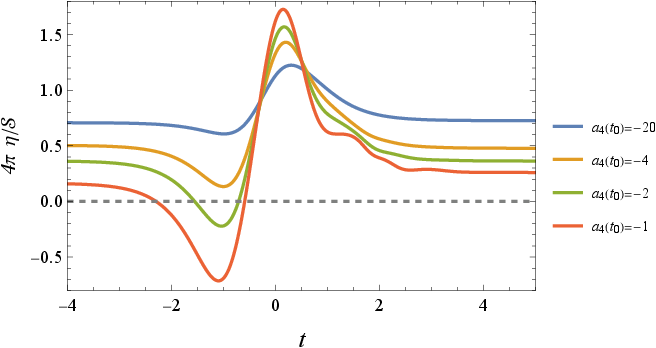}
    \caption{Shear viscosity to entropy density ratio, $\eta/\mathcal{S}$ (normalized to $1/4\pi$), in far-from-equilibrium isotropization with $c=2$. The lines with different colors correspond to different initial energy densities set by $a_4(t_0)$, see~\eqref{eqE}.}
    \label{fig:eta}
\end{figure}
The evolution of the viscosity-to-entropy ratio, $\eta / \mathcal{S}$, is shown in Figure~\ref{fig:eta} for different initial energy densities. The curves with smoother slopes correspond to larger $\mathcal{E}_0$.
Far from equilibrium, $\eta / \mathcal{S}$ changes significantly. When the boundary geometry undergoes rapid changes around $t = -1$, a notable decrease with a pronounced minimum is observed, followed by a rapid increase. The ratio decreases after reaching a maximum (around $t = 0$), with some cases (red curve) showing an oscillatory approach to stabilization at the final value. For small initial energy densities, $\eta / \mathcal{S}$ can become negative near $t = -1$ (green and red curves). In this regime, as shown in~\eqref{Etest}, the energy density decreases, corresponding to the energy drop observed in Figure~\ref{fig:Tuv}. However, the viscosity-to-entropy ratio remains positive for sufficiently large initial energy density (orange and blue curves). We highlight that the ratio attains a stable value at both early and late times, as the shear rate becomes too weak to affect the system in the limits $t \rightarrow \pm \infty$. Nonetheless, the early-time value of the ratio is lower than that at late times.

As a low-energy effective field theory for near-thermal equilibrium systems, fluid dynamics is formulated using a derivative expansion of local fluid mechanical variables. Up to the first velocity gradient, the shear viscosity, $\eta_0$, is given by the Navier-Stokes term in~\eqref{myeta0}. The apparent viscosity defined in~\eqref{dfeta} is expected to converge to the near-equilibrium viscosity, $\eta_0$, in the limit
\begin{equation}\label{condition}
 \dot{B}_0 \tau_c\ll 1\,,    
\end{equation}
where $\tau_c$ is the characteristic relaxation time of the system. However, as seen in Figure~\ref{fig:eta}, the late-time steady value of $\eta / \mathcal{S}$ is less than the KSS bound, $\eta_0 / \mathcal{S} = 1 / 4\pi\approx 0.80$, which was derived under equilibrium conditions using linear response theory. This discrepancy raises an important question: How can we reconcile the deviation from the KSS bound when the system appears to be in equilibrium at late times?

\begin{figure}[h]
    \centering
    \includegraphics[scale=1.0]{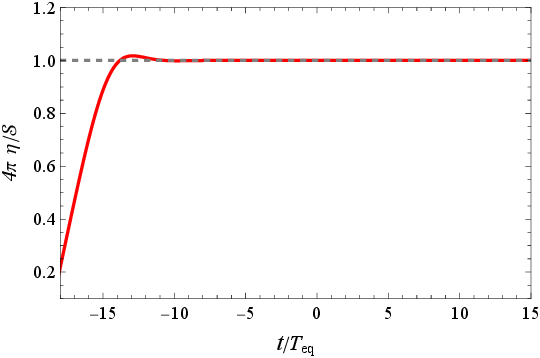}
    \caption{The evolution of $\eta/\mathcal{S}$ for $B_0(t)=\epsilon \sin(\omega t)$ with $\epsilon=0.0001$ and $\omega/T_{eq}=0.00031$. The evolution converges to first order viscous hydrodynamics very quickly and $\eta/s$ saturates at an exact value of $1/4\pi$ (dashed line).}
    \label{fig:eta_shy}
\end{figure}

One may wonder that in our present setup, the anisotropy cannot be attributed to hydrodynamic modes, as it appears no spatial momentum transfer and hydrodynamic modes are turned off. So what we propose to consider might be a fake shear viscosity that does not have anything to do with near equilibrium $\eta_0$ satisfying $\eta_0 / \mathcal{S} = 1 / 4\pi$ even at late times. To answer this point, we now show that our setup indeed recovers the standard near equilibrium shear viscosity $\eta_0$ when considering the form of shear deformation as the linear response approach \emph{i.e.} $B_0\sim e^{-i\omega t}$ with $\omega$ the frequency. In practice, we consider the shear flow with
\begin{equation}\label{Kobu}
  B_0(t)=\epsilon \sin(\omega\, t)\,,  
\end{equation}
where $\epsilon$ is a constant. For sufficiently small values of $\epsilon$ and $\omega$, one should recover the KSS result obtained from hydrodynamics. 
The viscosity-to-entropy ratio is presented in Figure~\ref{fig:eta_shy}. Apart from a short transient regime highly sensitive to the initial conditions, the ratio quickly approaches a
constant, which is given exactly by the close-to-equilibrium first-order viscous hydrodynamics. What's more, in the hydrodynamic limit for which $B_0=\epsilon e^{-i \omega t}$ and $\omega/T \ll 1$, one finds from~\eqref{newbk} that the shear deformation at linear order of $\epsilon$ becomes 
\begin{equation}
    ds^2 = -dt^2 + (dx^2 + dy^2 + dz^2) -2 \epsilon e^{-i \omega t} (dx \, dy + dy \, dz + dz \, dx)\,,
\end{equation}
and meanwhile 
\begin{equation} 
\delta T_{ij} =-i \omega\eta  B_0=-i \omega\eta  \delta g_{ij}\,,\quad i\neq j\,,\quad i, j=x,y, z\,,
\end{equation}
to the linear order in $\omega$.
This is what has been done in the holographic literature for computing the shear viscosity using standard linear response theory. Therefore, our setup gives a natural definition of shear viscosity beyond the regime of linear response theory (first-order viscous hydrodynamics).

While we cannot provide a complete explanation for the deviation from the KSS result at later times when~\eqref{condition} is stratified, a heuristic understanding is as follows. We first note that the condition in~\eqref{condition} does not guarantee the negligibility of higher-order derivative terms in hydrodynamics. More precisely, the higher-order derivative expansion typically includes two types of terms~\cite{Lublinsky:2009kv}: nonlinear terms in the fluid velocity, such as $(\nabla u)^2$ and $\sigma_{\mu\lambda}{\sigma^\lambda}_\nu$, and linear terms, like $\nabla \nabla u$ and $u^\lambda \nabla_\lambda \sigma_{\mu\nu}$. The nonlinear terms can be neglected when the amplitudes of local fluid mechanical variables are small (\emph{e.g.} in our case, $\dot{B}_0 \rightarrow 0$). However, even for small amplitude perturbations, significant contributions can arise from the linear terms when the momenta associated with fluid perturbations are large, which is beyond the first order viscous hydrodynamics. Therefore, in addition to~\eqref{condition}, to neglect higher-order derivative linear terms, we should require
\begin{equation}\label{condition2}
\tau_c^{n-1} \partial^n_t B_0\ll \dot{B}_0\,,  
\end{equation}
where $n$ is a positive integer. The smaller the value of $n$ is, the higher-order terms can be neglected. For example, near equilibrium, the viscosity $\eta_0$ is extracted using the standard Kubo formula with the perturbation $B_0 \sim e^{-i \omega t}$ by taking the limit $\omega/T \ll 1$. It is clear that~\eqref{condition2} is satisfied in this case, for which one should have the KSS result $\eta_0 / \mathcal{S} = 1 / 4\pi$ in the hydrodynamic limit (as confirmed in Figure~\ref{fig:eta_shy}).

However, in the present case with the shear deformation of~\eqref{eqb0}, while the first condition in~\eqref{condition} is fulfilled at late times, the second one in~\eqref{condition2} cannot be satisfied. For example, we find that $\lim_{t \rightarrow \infty} \frac{\ddot{B}_0}{\dot{B}_0} = -2$ and $\lim_{t \rightarrow \infty} \frac{\dddot{B}_0}{\dot{B}_0} = 4$. More generally, the higher the derivative considered, the larger this ratio becomes. Consequently, the linear response formalism and hydrodynamic approximation are no longer applicable, and the KSS result is not recovered at late times.\footnote{It was found that, when the shear rate in units of the energy density is small at late times, the viscosity-to-entropy density ratio in far-from-equilibrium strongly coupled fluids coincides with the near-equilibrium hydrodynamic expectation~\cite{Baggioli:2021tzr}, where both~\eqref{condition} and~\eqref{condition2} are satisfied.} In this sense, the definition of $\eta$ in~\eqref{dfeta} accounts for both nonlinear and non-hydrodynamic mode contributions. Thus, the holographic framework allows for a resummation of the hydrodynamic description, extending it beyond the regime where higher-order terms contribute comparably to the lowest-order terms~\cite{Heller:2013fn}. Our findings are consistent with~\cite{Lublinsky:2009kv}, where the authors introduced an effective shear viscosity by including higher-order derivative terms and showed that these terms tend to reduce the viscosity’s impact. Moreover, it was argued that the higher order dissipative terms strongly reduce the effect of the usual viscosity and an effective viscosity-to-entropy ratio found from comparison of Navier-Stokes results to experiment, can be below the KSS bound~\cite{Lublinsky:2007mm}.
Indeed, the Borel-resummed, out-of-equilibrium shear viscosity defined by~\eqref{dfeta} has been shown to approach zero for far-from-equilibrium systems~\cite{Romatschke:2017vte} (see also~\cite{Baggioli:2021tzr}). 

\begin{figure}[h]
    \centering
    \includegraphics[scale=0.8]{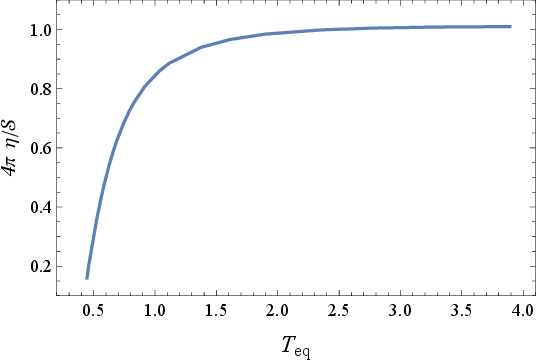}
    \includegraphics[scale=0.8]{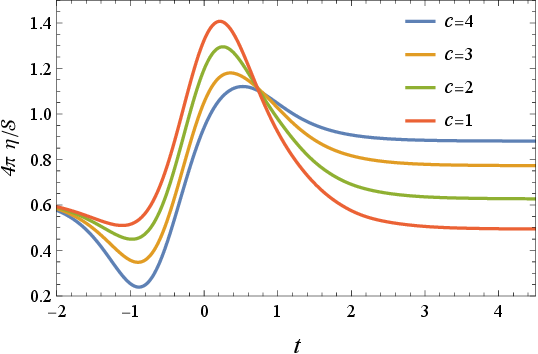}  
    \caption{\textbf{Left}: The late time equilibrium $\eta/\mathcal{S}$ as a function of the equilibrium temperature $T_{eq}$ for $c=2$. It approaches $1/4\pi$ from below as the temperature is increased. \textbf{Right}: The evolution of $\eta / \mathcal{S}$ during far-from-equilibrium isotropization for different values of $c$ with the same initial data.}
    \label{fig:etavsc}
\end{figure}
The temperature gives the only reasonable microscopic length scale in our neutral, conformal, strongly coupled fluids. As the intrinsic micro-scale, $1/T$, decreases with increasing temperature, dissipative effects from higher-order terms are expected to diminish, leading to the KSS result. The equilibrium value of $\eta / \mathcal{S}$ at late times is shown as a function of $T_{\text{eq}}$ in the left panel of Figure~\ref{fig:etavsc}. We find that $\eta /\mathcal{S}$ approaches $1/4\pi$ from below in the high-temperature limit, where higher-order terms are suppressed.  Conversely, the viscosity can be parametrically smaller than the KSS value by lowering $T_{\text{eq}}$ (or equivalently by choosing initial data with a small energy density). Additionally, the final equilibrium temperature, $T_{\text{eq}}$, increases as the quench strength $c$ of~\eqref{eqb0} is increased. Therefore, we expect that $\eta / \mathcal{S}$ at late times will approach $1/4\pi$ by increasing $c$ while keeping the initial data fixed. This trend is confirmed in the right panel of Figure~\ref{fig:etavsc}.

\begin{figure}[h]
    \centering
    \includegraphics[scale=1.0]{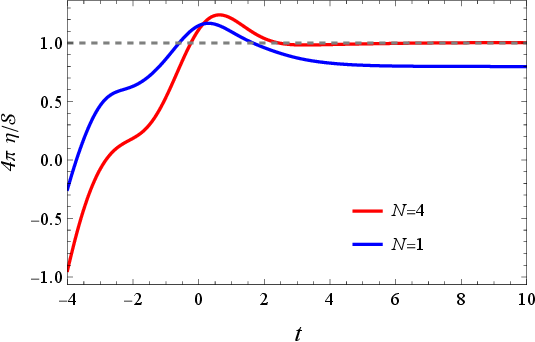}
    \caption{The evolution of $\eta/\mathcal{S}$ in far-from-equilibrium isotropization for the new quench function~\eqref{newB}. The red line is for $N=4$, and the blue one is for $N=1$. The former case saturates at a value very close to $1/4\pi$ while for the latter $\eta/\mathcal{S}\approx 0.063$ at late times. We choose a small value of initial energy density with $c=2$.}
    \label{fig:newB}
\end{figure}

We have argued that the condition of~\eqref{condition2} is necessary for the hydrodynamic limit, ensuring that higher-order derivative terms are negligible. Thus, if we select a new quench function $B_0(t)$ such that~\eqref{condition2} is satisfied at late times, we should recover the KSS result as the system asymptotically becomes static. We take
\begin{equation}\label{newB}
    B_{0}(t) = \frac{c}{(1 + e^{t / \tau})^{1 / N}} \,,
\end{equation}
where $N$ is a positive constant. Furthermore, we set $\tau = 1$ due to the absence of other scales in conformally invariant supersymmetric Yang-Mills theory. It follows that
\begin{equation}
    \lim_{t \rightarrow \infty} \Bigg| \frac{B_0^{(n)}}{\dot{B}_0} \Bigg| = \left( \frac{1}{N} \right)^{n - 1} \,.
\end{equation}
Thus, when $N > 1$, higher-order derivative terms decay rapidly as $n$ increases. As shown in Figure~\ref{fig:newB}, the viscosity-to-entropy ratio for $N = 4$ (red curve) saturates at a value very close to $1 / 4\pi$, even with a small initial energy density. The critical case with $N = 1$ (blue curve) is provided for comparison, where the ratio saturates at a value well below $1 / 4\pi$.

\section{Conclusion and Perspective}\label{sec:conc}

A particularly useful and widely studied model of QCD plasma is $\mathcal{N}=4$ supersymmetric plasma through the AdS/CFT correspondence. We have investigated the isotropization dynamics of supersymmetric Yang-Mills plasma in far-from-equilibrium scenarios for which the energy-momentum tensor expectation value at the conformal boundary can be extracted explicitly.
As shown in Figure~\ref{fig:b4}, irrespective of significant initial anisotropy, the system quickly converges towards an attractor driven by the decay of non-hydrodynamic modes. The evolution of the energy density, transverse pressure, and longitudinal pressure was tracked throughout the far-from-equilibrium process, revealing that the isotropization time depends on both the strength of the applied quench and the system's initial energy density (see Figure~\ref{fig:tau}). The large initial energy densities significantly accelerate the isotropization, which breaks the isotropization behavior in previous work~\cite{Chesler:2008hg}. Nevertheless, the entropy production~\eqref{eq:ds} is always positive. 

We have introduced an effective shear viscosity beyond the regime of linear response theory. Note, however, that our shear viscosity agrees exactly with the one from the Kubo formula when taking the hydrodynamics limit with shear deformation $\sim e^{-i \omega t}$ and $\omega/T \ll 1$, see Figure~\ref{fig:eta_shy}. We have found that the shear viscosity to entropy density ratio ($\eta/\mathcal{S}$) exhibits significant variations during the evolution, deviating from the near-equilibrium KSS bound, see Figure~\ref{fig:eta}. The numerical results showed that non-equilibrium shear flows have rich dynamics, illustrating how $\eta/\mathcal{S}$ evolves in response to boundary shear rates and initial energy densities. Interestingly, we have found that the late-time equilibrium value of $\eta / \mathcal{S}$ is generally less than the KSS bound, $\eta_0 / \mathcal{S} = 1 / 4\pi$.\,\footnote{We highlight that the late-time equilibrium $\eta / \mathcal{S}$ of Figure~\ref{fig:eta} is beyond the description of the first-order viscous hydrodynamics. The latter gives the KSS result. } Nevertheless, the value of such equilibrium ratio increases as both the strength of quench and initial energy density (or equivalently $T_{eq}$ ) are increased, approaching the KSS bound $1/4\pi$ from below (see Figure~\ref{fig:etavsc}). We have provided a heuristic understanding by noting that our effective viscosity coefficient $\eta$ accounts for both nonlinear and non-hydrodynamic mode contributions and allows for a resummation of the hydrodynamic description. However, a complete understanding is still lacking. Based on our picture, the equilibrium value of $\eta / \mathcal{S}$ should depend on the driving function. In particular, in cases where higher derivative terms are negligible at late times, one will recover the KSS result. This was demonstrated by a new quench function~\eqref{newB}, see the red curve of Figure~\ref{fig:newB}. Given that the shear viscosity is typically measured in the lab via~\eqref{dfeta}. One should be careful when comparing the measurement results to the theoretical one from the Kubo formula corresponding to the particular deformation form in~\eqref{Kobu}.

Our findings align with the modern understanding of hydrodynamization while highlighting novel features of far-from-equilibrium shear transport. Our approach complements recent studies of hydrodynamic attractors~\cite{Heller:2015dha, Kurkela:2018vqr} and provides a template of isotropization for extending holographic models to QCD-like plasmas. This underscores the effectiveness of holographic models in capturing the non-equilibrium behavior of strongly coupled systems like QCD and highlights the need for non-linear and non-hydrodynamic approaches under strong quenches. One pressing direction of research
is to understand the extent to which these are useful results for experimental systems involving QCD plasma. Incorporating finite baryon density into the analysis could provide insights into how baryon number influences isotropization times and the overall dynamics of QCD matter. Future studies could extend the current analysis to include external fields, such as magnetic or rotational effects. In the present study, we have limited ourselves to the toy mode, \emph{i.e.} $\mathcal{N}=4$ supersymmetric Yang-Mills plasma. In previous works by some of us, quantitative holographic QCD models were constructed for pure gloun~\cite{He:2022amv}, 2-flavor QCD~\cite{Zhao:2023gur} and (2+1)-flavor QCD~\cite{Cai:2022omk,Cai:2024eqa}. It is interesting to generalize our study to these holographic models that quantitatively describe the equation of state of QCD. These directions will deepen our understanding of the real-time dynamics of QCD in the presence of complex external conditions, bridging the gap between theoretical predictions and experimental observations.

\acknowledgments

We thank Jia Du, Yuan-Xu Wang, and Haotian Sun for their helpful discussions. This work is supported in part by the National Natural Science Foundation of China Grants No.12075298, No.12075101, No.12475053, No.12122513, No.12347209, No.12047569, No.12235016, and No.12447101. SCW is supported by the Fellowship of China Postdoctoral Science Foundation No.2022M713227. S.H. would also like to express appreciation for the financial support from the Max Planck Partner Group. We acknowledge the use of the High Performance Cluster at the Institute of Theoretical Physics, Chinese Academy of Sciences.

\bibliographystyle{JHEP}
\bibliography{reference}

\end{document}